\documentclass[sigconf,nonacm]{acmart}
\usepackage[utf8]{inputenc}
\usepackage{multirow}

\AtBeginDocument{%
  \providecommand\BibTeX{{%
    \normalfont B\kern-0.5em{\scshape i\kern-0.25em b}\kern-0.8em\TeX}}}

\copyrightyear{2020} 
\acmYear{2020} 
\setcopyright{acmlicensed}\acmConference[UMAP '20]{Proceedings of the 28th ACM Conference on User Modeling, Adaptation and Personalization}{July 14--17, 2020}{Genoa, Italy}
\acmBooktitle{Proceedings of the 28th ACM Conference on User Modeling, Adaptation and Personalization (UMAP '20), July 14--17, 2020, Genoa, Italy}
\acmPrice{15.00}
\acmDOI{10.1145/3340631.3394873}
\acmISBN{978-1-4503-6861-2/20/07}

\begin{document}

\title{More Than Accuracy: Towards Trustworthy Machine Learning Interfaces for Object Recognition}


\author{Hendrik Heuer, Andreas Breiter}
\affiliation{%
  \institution{University of Bremen, Institute for Information Management, Germany}
}
\email{hheuer@ifib.de, abreiter@ifib.de}

\renewcommand{\shortauthors}{Heuer and Breiter}

\begin{abstract}
This paper investigates the user experience of visualizations of a machine learning (ML) system that recognizes objects in images. This is important since even good systems can fail in unexpected ways as misclassifications on photo-sharing websites showed. In our study, we exposed users with a background in ML to three visualizations of three systems with different levels of accuracy. In interviews, we explored how the visualization helped users assess the accuracy of systems in use and how the visualization and the accuracy of the system affected trust and reliance. We found that participants do not only focus on accuracy when assessing ML systems. They also take the perceived plausibility and severity of misclassification into account and prefer seeing the probability of predictions. Semantically plausible errors are judged as less severe than errors that are implausible, which means that system accuracy could be communicated through the types of errors.
\end{abstract}



\keywords{User Experience; Human-Centered Machine Learning; Algorithmic Experience; Algorithmic Bias; Trust;}

\maketitle

\section{Introduction}

Machine learning (ML) systems are designed to help users manage their photo collections, provide image captions for blind and visually impaired people, and enable self-driving cars to detect traffic signs and pedestrians~\cite{google_photos_2015,MacLeod:2017:UBP:3025453.3025814,6301978,bojarski2016end}. An HCI perspective on how to expose such ML systems that learn from data to users is only emerging~\cite{hardt_how_2014,STUMPF2009639,kim2015interactive,Dove:2017:UDI:3025453.3025739,Veale:2018:FAD:3173574.3174014}. In this paper, we explore what visualizations need in order to be perceived as transparent and trustworthy, which visualizations support the users in assessing the performance of object recognition systems and how different visualizations influence a user's trust. Our goal is to examine the user experience of simple visualizations and their influence on trust. Trust is an important way to cope with risk, complexity, and a lack of system understanding~\cite{luhmann_trust_1979}. Since ML systems can be complex, hard to understand and prone to fail at unexpected times, it is important to understand what makes peoples trust such systems. While a large body of research is already focused on developing complex visualizations of machine learning systems~\cite{DBLP:journals/corr/YosinskiCNFL15,DBLP:journals/corr/SamekBMBM15,DBLP:journals/corr/StrobeltGHPR16,7539329}, a thorough investigation of the user experience of ML systems in use is missing. In this paper, we examine how expert users assess the performance of a system based on the output of the system itself. ML systems often do not have dedicated interfaces. Users are usually just presented with the output of an ML system, often without any indication that they are interacting with an ML system. In this study, we contribute to the understanding of the user experience needs of such ML systems. 

We want to support users with a basic understanding of machine learning systems in assessing a system in use. We investigated the visualizations and systems with experts from two groups: students who took an ML class, and engineers who develop ML systems. The participants in our study interacted with an object recognition system that labels images into ten different categories. Participants reviewed ten predictions for each of the 3x3 combinations of visualizations and machine learning systems. The ML systems achieve different accuracy, the visualizations differ in their complexity and in how much of an ML system they expose. After the participants rated the systems, we interviewed them about their experience. We answer the following research questions: 

\begin{itemize}
\item Which of the visualizations help users with a background in ML assess the performance of ML systems?~(RQ1)
\item What makes a visualization helpful for the assessment of ML systems?~(RQ2)
\item How do the visualization and the accuracy of an ML system affect users' trust and reliance?~(RQ3)
\end{itemize}

\section{Related Work}

The prevalence of complex, opaque, and invisible algorithms that learn from experience and data motivated a variety of investigations of algorithm awareness and algorithmic bias~\cite{Hamilton:2014:PUE:2559206.2578883,eslami_i_2015}. Algorithmic bias, whether unintentional or intentional, was found to severely limit the perceived performance of an ML system. This motivated us to explore whether participants can assess the accuracy of an object recognition system in use, i.e. based on the system's predictions. Our goal is to develop interfaces that support users in assessing the accuracy of an object recognition system in use.

To support trust in ML systems, we investigated different interfaces that displayed visualizations. Visualizations are a common way of communicating algorithmic processes~\cite{gregor1999explanations,kim2015interactive,STUMPF2009639}. A good visualization communicates complex ideas with clarity, precision, and efficiency~\cite{Tufte:1986:VDQ:33404}. We extend on prior work on how end-users interact with machine learning systems~\cite{amershi_power_2014,STUMPF2009639,kim2015interactive}. The research question about the helpfulness of visualization in assessing the accuracy of an ML system in use connects to Rader et al., who explored different explanation styles to communicate the news curation system on Facebook. They showed that explanations can help users determine whether a system is biased~\cite{Rader:2018:EMS:3173574.3173677}. We focus on the user experience of simple visualizations and whether users can assess the accuracy of an ML system based on its output. Examples for visualizations of neural networks include showing the decision function~\cite{DBLP:journals/corr/SelvarajuDVCPB16}, visualizing the filters~\cite{DBLP:journals/corr/SpringenbergDBR14} or providing explanations~\cite{Ribeiro_2016_WIT}. Such visualizations are highly specialized and presented without an empirical basis for their helpfulness~\cite{DBLP:journals/corr/YosinskiCNFL15,DBLP:journals/corr/SamekBMBM15,DBLP:journals/corr/StrobeltGHPR16,7539329}. The large majority of ML papers does not evaluate their visualizations in user tests and does not compare their visualizations to basic visualizations~\cite{Ribeiro_2016_WIT,DBLP:journals/corr/abs-1804-11191}.

In this paper, we focus on trust in intelligent systems~\cite{tullio2007works,herlocker2000explaining} and automation~\cite{muir_trust_1994,muir_trust_1996,lee2004trust} in the context of object recognition. This context poses many risks that range from a cluttered photo collection to blind people becoming misinformed and self-driving cars hurting people. These risks directly connect to the complexity of deep neural networks, which are employed for many tasks. Even ML experts cannot fully understand or explain how and why deep neural networks make individual decisions. Since a full understanding is not possible, trust is important. This connects to prior work by Lee and See, who show that whether a task like deleting spam email is delegated depends on the trust in a system \cite{lee2004trust}. They state that trust guides reliance when the complexity of the automation makes a complete understanding impractical, which directly relates to the black box scenario in machine learning. Prior work by Muir~\cite{muir_trust_1994}, who modeled trust in a machine based on interpersonal trust, showed that users can meaningfully rate their trust in intelligent systems. Cramer et al. adopted Muir's definition in the context of spam filters and showed that awareness and understanding seriously impact the users' attitudes and behavior~\cite{cramer_awareness_2009}. 

\section{Method}

\begin{figure}
\centering
  \includegraphics[width=\columnwidth]{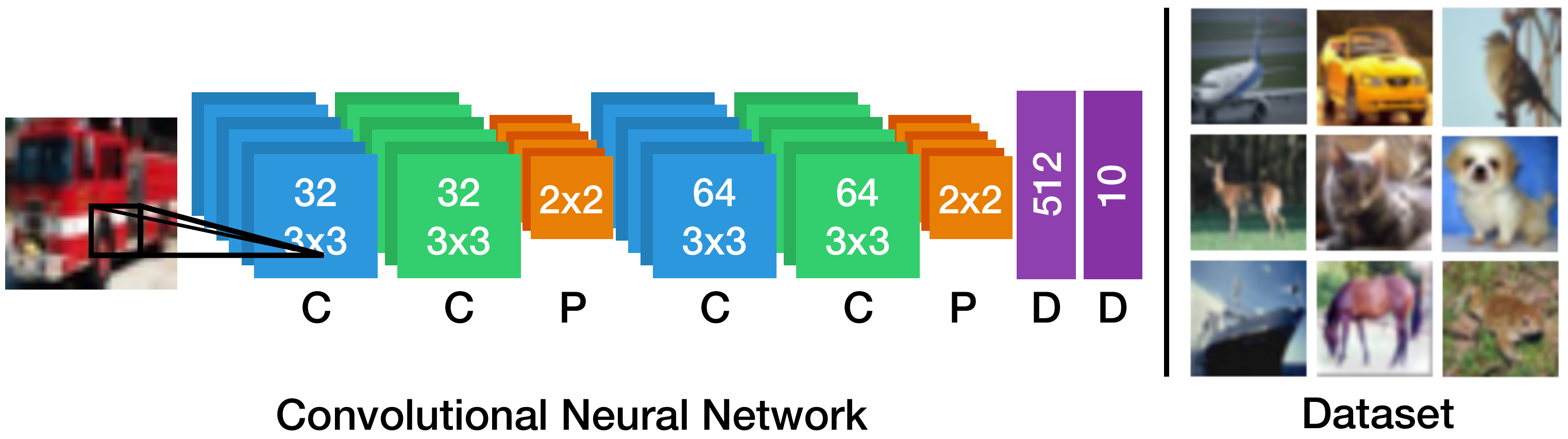}
  \caption{We trained three convolutional neural networks on the CIFAR-10 object recognition dataset with ten classes.}~\label{fig:machine_learning_system}
\end{figure}

\begin{figure}
\centering
  \includegraphics[width=.8\columnwidth]{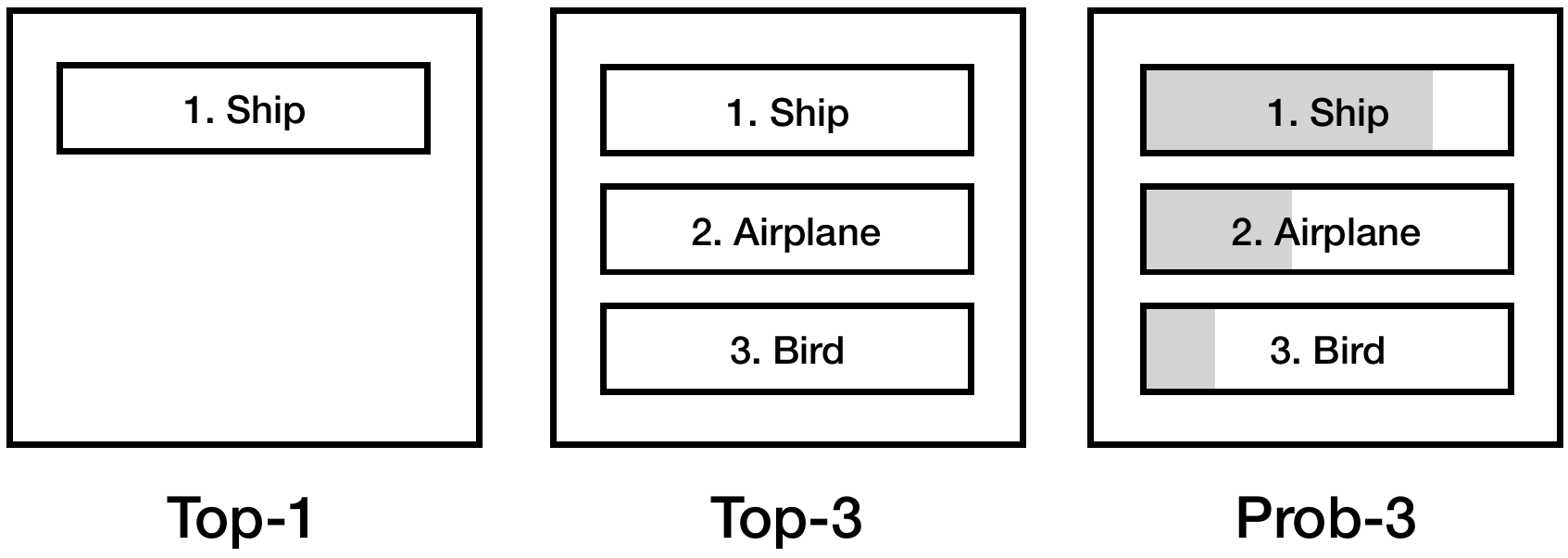}
  \caption{The three visualizations evaluated in this study.}~\label{fig:visualizations}
\end{figure}

For this study, we interviewed expert users after they reviewed ML-based object recognition systems. In a within-subject study, each participant rated 3x3 combinations of visualizations and machine learning systems via a web application. The ML systems had different levels of accuracy on an unseen test set. Participants were shown one image and one visualization at a time. After a waiting period of five seconds, participants also classified each of the images themselves. The waiting period was added to force participants to look at each image and each prediction carefully and keep them from clicking without looking. The order of the images was randomized but the same for each participant.

The study was conducted with machine learning systems trained to classify images based on objects. The ML systems were trained on the CIFAR-10 dataset, which consists of 60,000 32x32 color images and the ten classes: airplane, bird, car, cat, deer, dog, frog, horse, ship, and truck~\cite{krizhevsky2009learning}. The CIFAR-10 dataset provides a task that is realistic, of limited scope (only ten classes), and so unique that none of the participants had prior experience in analyzing the output of such a system~\cite{DBLP:journals/corr/Graham14a,DBLP:journals/corr/MishkinM15, liang2015recurrent}. To compare systems with different levels of accuracy, we trained three nine-layer convolutional neural networks for two, ten and 160 epochs. We refer to the three systems we trained as Bad, Medium, and Good. The names are based on how many errors the systems make on average. The Bad system was trained for two epochs and has an accuracy of 51\% (for the unseen testing data). The Medium system was trained for ten epochs and reached an accuracy of 65\%. The Good system was trained for 160 epochs and achieved an accuracy of 83\%. Training three neural networks with different levels of accuracy is challenging since the accuracy of an ML system does not improve linearly. Participants were presented with the three visualizations shown in Figure~\ref{fig:visualizations}: Top-1, Top-3, and Prob-3. The Top-1 visualization shows the prediction with the highest probability. This visualization was selected because it minimizes complexity and cognitive load by giving a simple, action-oriented answer to the classification task. It is the most commonly used interface for object recognition systems like Google Photos and Flickr~\cite{hern_flickr_2015,ahmed_google_2015}. Top-3 shows the top three predictions ranked by their probability. We hypothesized that Top-3 would allow participants to make a more informed decision about the ML system since participants can use the ranking to compare the predictions. Prob-3 shows the same ranking and visualizes the probability of each prediction using a bar. The probabilities are based on the activation of the softmax unit of the convolutional neural network, which represents the probability distribution over ten different classes~\cite{Goodfellow-et-al-2016}. A high activation of a particular neuron indicates a high probability that this particular class is the correct classification of the image. Prob-3 was selected since it visualizes the probability distribution computed by the machine learning system, thus exposing that such ML systems operate in the regime of probabilities and correlations.

We investigate which visualizations help users with a background in ML to assess the performance of ML systems. We want to support people in assessing the trustworthiness of systems in use. Therefore, we conducted our investigation with a sample of experts that had sufficient knowledge of ML. We recruited our experts from two groups: ML students and ML engineers. Fifteen~ML students~(13 male, two female) were recruited at a large campus university in Germany. All students had attended a machine learning or data science course. The five ML engineers (all male) were recruited from a German augmented reality company with an ML focus. The study and the interviews were conducted in German and translated into English before the coding. Informed by prior work that focused on extreme users to gain rich insights into specific issues like customization in communication apps~\cite{griggio_customizations,Choe:2014:UQP:2556288.2557372,Djajadiningrat:2000:IRE:347642.347664}, we conducted an investigation with ML experts and students. This serves two purposes: (1) it allows us to investigate the needs that experts have when evaluating ML interfaces for object recognition, and (2), by extension, it allows us to understand how ML-based interface for photo-sharing websites or tools to manage photo collections need to be designed for end-users without a background in ML. 

In the interviews, we elicited how the participants assess the accuracy of the ML systems and what information they believe they need for the assessment. We interviewed 15 participants (13 male, two female), who all had used the 3x3 combinations of visualizations and ML systems before the interview. Of the 15 interviewees, ten were students and five were engineers. We will cite statements by the interviewees as follows: If the fifth student made a statement, we indicate this as~S5, while the third engineer is cited as~E3. The interviews were audio-recorded and transcribed. In the semi-structured interviews, participants were asked how they determined the accuracy of a system, which visualization they found most helpful, and which visualization they found least helpful. We also asked participants to provide reasons for their answers. In addition to that, participants were asked whether they distinguished between ``relying on the prediction of the system'' and ``trust in the system'', which was followed by the question ``If so, what was the difference for you?''. This question was motivated by trust definitions that equal trust to reliance~\cite{rousseau_not_1998,lee2004trust}. Finally, participants were asked the open question: ``What does trust mean to you in terms of machine learning?''. We coded the qualitative data using an inductive approach, where we read the raw data in detail and clustered textual quotations to identify emergent themes~\cite{mayring2007generalization}. The material was analyzed step-by-step, assigning codes to individual words and sentences. After that, the material was revisited based on the codes that emerged. All codes were collected, sorted, and then combined into inductive categories. These analytical units were then sorted by the number of codes in the category and combined into inductive categories with a higher level of abstraction, merging semantically similar categories.

\section{Results}

The first research question was which visualizations help users assess the performance of ML systems. We asked participants to name the most and the least helpful visualization. Of the three visualizations, all participants considered Prob-3 to be the most helpful visualization. Top-1 and Top-3 were never mentioned as the most helpful visualization. Top-1 was considered the least helpful visualization by the majority~(66\%) of people: three engineers and seven students named it in the interviews. Five participants (three students, two engineers) considered Top-3 to be the least helpful, though one participant also said that he found none of the visualizations helpful to assess the accuracy of the systems~[E1].

Next, we investigated what makes a visualization helpful for this assessment~(RQ2). For this, we explored the criteria and strategies that the participants apply to assess the ML visualizations. We found that participants want visualizations to expose the inner mechanics of the system~[S2]. They highlighted the importance of more information~[S4, S5], which allowed them to rely on the system to produce accurate results~[S5] or since any information helps~[E4]. One participant stated that the Prob-3 visualization showed that the system does not have a crystal ball that tells the truth, but that it is ``all about probabilities''~[S10]. However, he also commented on the diminishing returns of additional information or situations of too much information~[S10]. One participant commented on the necessary compromise between clarity and the amount of information displayed~[S1]. 

A large group of participants stated that they use the probability of a prediction and its visualization as an indicator to make their decision~[E2, E3, E5, S1, S10]. The display of probabilities was seen as a big improvement~[S1]. People noted that seeing the probability makes it possible to decide whether something has a high probability or whether a prediction is just due to chance~[S8]. Not knowing about the probability distribution was considered to be a problem with some of the visualizations~[S8, S9]. Besides looking at the probabilities, counting errors or counting correct classifications were commonly cited ways to assess the accuracy of an ML system~[E1, S1, S5, S7, S9]. Participants also used the frequency of the errors~[E1] and their severity~[S3]. Some participants explicitly framed this as comparing the system accuracy to their accuracy at the task~[E4, S5, S7]. For one, errors confirmed that the system has trouble with the prediction task~[S10]. Some participants found it hard to determine the accuracy of the system and commented on the difficulty and complexity of the task~[E1, S4, S10]. One participant remarked that the visualizations only provided a snapshot of the accuracy of the system, which led him to still have doubts about the overall accuracy of the system~[S3]. Another participant commented that the speed of the assessment of the system performance is most important to him~[E5]. A large group also viewed it positively when the wrong predictions were at least plausible for them as humans~[E5, S1, S3, S6, S9]. Animals, for example, should yield other animals~[E5, S1, S6]. Mistaking a cat for a dog was considered a less severe error~[S6] and better than mistaking a cat for an airplane~[S1]. A complete mismatch between the predicted classes and the true classes led to the system being perceived as being incorrect and faulty~[S9]. For situations where more than one prediction was shown, the cognitive load was considered higher since the user had to reason about the semantic similarity of the different categories~[S9]. If a visualization provided a ranking, participants used their decisions based on whether the top prediction was correct~[S8]. For a visualization that also showed the probability of the decision respectively the confidence of the system, the criterion was whether the probability was higher than 50 percent~[S8].

The third research question was how the visualizations and the accuracy of the ML system affect trust and reliance. In the following, we will, therefore, explore how the participants differentiated between trust and reliance in the interviews. A large fraction of our participants equaled trust to accuracy~[E2, E5, S1, S2, S4, S7, S10]. They saw trust to be based on observing a mismatch between the prediction of the system and their classification of the image~[E1], a function of accuracy~[E5], or somehow connected to accuracy~[S5]. For one participant, the expectation that the results are correct equals trust~[S4]. Other participants framed trust as high prediction accuracy~[S7] or focused on a negative relationship where less accuracy equals less trust~[E3, S1, S7]. The accuracy metric was seen as a way to build trust~[S5]. Trust was described as giving up agency and letting the system make decisions on behalf of the user~[S10]. One participant connected trust to feelings~[S5]. This emotional perspective was also voiced by another participant, who connected trust to feelings and intuition~[E4]. One participant described a moment where an error caused a breakdown that eroded his trust: ``Suddenly, the moment that I found the first error in the system, I was like Oh! (…) it destroyed my trust. And after that, I could not trust it''~[E3]. Other participants commented on how trust develops and improves through the understanding of a process~[S2] and with time~[S5, S6]. Participants also commented on the connection between experience and trust~[S6], that they have more trust in an existing system than in a new system~[S5], and that knowing a system influences both trust and reliance~[S5]. When asked to distinguish between relying on the predictions of a system and trusting a system, many participants regarded trust and reliance in a system as the same~[E2, S4, S7, S8], mostly the same~[S10], or similar~[S1, S2]. Three participants made a difference between trust and reliance~[E5, S3, S5]. They stated that in some cases, trust is not the same as reliance~[S5], even though the difference is not clear~[E5]. Regarding the relationship between trust and reliance, one participant said that when ``I rely on it, I trust it''~[S4]. Another participant said that he would rely on a system, but he would not trust it~[S6]. Others said that trust comes first and reliance second~[E3, S8]. One of the participants regarded trust as a necessary condition for reliance~[S8]. Reliance was also described as not checking again whether something is correct~[S5]. One participant made the distinction of reliance as technical and trust as emotional~[S6]. 

\section{Discussion}

The results show that participants prefer visualizations that expose the statistical nature of ML systems by indicating the probability distribution of the Top-3 predictions over action-oriented visualizations that only show the top prediction. This has important implications for interactive machine learning and applications where users are directly and indirectly training machine learning systems. Such systems should expose the probabilities computed by the machine learning systems. With this, we extend on recent research that studies algorithmic experience and awareness~\cite{eslami_i_2015,Hamilton:2014:PUE:2559206.2578883,Alvarado:2018:TAE:3173574.3173860}. Our results also connect to Eslami et al.'s finding that users can detect algorithmic bias during their regular usage of a service~\cite{eslami2017careful}.

A trustworthy visualization for an object recognition system should support the strategies that participants apply to distinguish between the systems. The most important strategies to determine the accuracy were interpreting the probabilities and counting the errors. For both, participants took the severity of errors into account. Regarding the helpfulness of visualizations, participants frequently commented on the appropriate amount of information. In the interviews, trust and reliance were strongly linked to accuracy. Further research should explore whether directly rating the accuracy of an ML system is the most straightforward way to measure the quality of an ML system or whether experiential qualities like trust and reliance capture more than mere accuracy. Our interviews showed that participants not only focus on accuracy but that they also take the perceived plausibility and severity of misclassification into account. It would, therefore, be interesting to see how ratings of trust extend on ratings of accuracy by accounting for the unexpected mistakes that ML systems can make. Our findings show that it may not be sufficient to merely improve the accuracy of an ML system. In the interviews, participants argued that mistaking a cat for a dog was better than mistaking a cat for an airplane~[S6] and that animals should yield other animals~[E5]. This implies that a semantically plausible error is regarded as more tolerable than a semantically unrelated error. Whether machine learning systems that fail in ways that humans consider semantically plausible are preferable remains an important open question. Our results indicate that the kinds of errors an ML system makes could be an important way to increase trust in an ML system. Optimizing the plausibility of errors and designing the error cases could be an alternative to explanations of the decisions of machine learning systems. This connects to Gregor et al., who argue that explanations are needed when unexpected behavior like anomalies occur~\cite{gregor1999explanations}. Anomalies, by definition, are hard to anticipate or to understand. Trust is aimed at situations where a complete understanding is impractical~\cite{luhmann_trust_1979,cramer_awareness_2009}. Rather than aiming to explain how decisions are made~\cite{DBLP:journals/corr/SelvarajuDVCPB16,DBLP:journals/corr/SpringenbergDBR14,Ribeiro_2016_WIT}, the plausibility of the ML predictions could be a straightforward way to communicate to users that they can trust and rely on a system in use. The types of errors a machine learning system makes could be designed to signal system accuracy. Plausible errors could increase trust in an ML system, while implausible errors could lead to distrust. The objective function of a neural network could be adapted to penalize errors that users regard as plausible and reward errors that users perceive as implausible. This could convey the system accuracy through the predictions of the machine learning system~\cite{Goodfellow-et-al-2016}. We also found that visualizations that show the probability of the predictions are preferred over rankings of the results. The interviews revealed the strategies that participants use to assess the performance of ML systems and what participants think they need in a visualization. When assessing ML systems that classify images, participants take the severity of an error and its likelihood into account. Semantically plausible errors are more tolerable than those that are seemingly random. We also found that the participants in our investigation did not want simple answers or recommendations they can follow. In conclusion, this means that trustworthy machine learning interfaces should show an appropriate amount of information, account for unexpected errors, and expose how the system works. Errors of the system could be used to communicate system performance. Semantically implausible errors, like a cat that is misclassified as an airplane, could be used to alert users about situations where a system is not trustworthy. 

\bibliographystyle{ACM-Reference-Format}
\bibliography{references}


\begin{thebibliography}{44}


\ifx \showCODEN    \undefined \def \showCODEN     #1{\unskip}     \fi
\ifx \showDOI      \undefined \def \showDOI       #1{#1}\fi
\ifx \showISBNx    \undefined \def \showISBNx     #1{\unskip}     \fi
\ifx \showISBNxiii \undefined \def \showISBNxiii  #1{\unskip}     \fi
\ifx \showISSN     \undefined \def \showISSN      #1{\unskip}     \fi
\ifx \showLCCN     \undefined \def \showLCCN      #1{\unskip}     \fi
\ifx \shownote     \undefined \def \shownote      #1{#1}          \fi
\ifx \showarticletitle \undefined \def \showarticletitle #1{#1}   \fi
\ifx \showURL      \undefined \def \showURL       {\relax}        \fi
\providecommand\bibfield[2]{#2}
\providecommand\bibinfo[2]{#2}
\providecommand\natexlab[1]{#1}
\providecommand\showeprint[2][]{arXiv:#2}

\bibitem[\protect\citeauthoryear{Ahmed}{Ahmed}{2015}]%
        {ahmed_google_2015}
\bibfield{author}{\bibinfo{person}{Kamal Ahmed}.}
  \bibinfo{year}{2015}\natexlab{}.
\newblock \showarticletitle{Google apologises for racist blunder}.
\newblock \bibinfo{journal}{\emph{BBC News}} (\bibinfo{date}{July}
  \bibinfo{year}{2015}).
\newblock
\urldef\tempurl%
\url{https://www.bbc.co.uk/news/technology-33347866}
\showURL{%
\tempurl}


\bibitem[\protect\citeauthoryear{Alvarado and Waern}{Alvarado and
  Waern}{2018}]%
        {Alvarado:2018:TAE:3173574.3173860}
\bibfield{author}{\bibinfo{person}{Oscar Alvarado} {and}
  \bibinfo{person}{Annika Waern}.} \bibinfo{year}{2018}\natexlab{}.
\newblock \showarticletitle{Towards Algorithmic Experience: Initial Efforts for
  Social Media Contexts}. In \bibinfo{booktitle}{\emph{Proceedings of the 2018
  CHI Conference on Human Factors in Computing Systems}}
  \emph{(\bibinfo{series}{CHI '18})}. \bibinfo{publisher}{ACM},
  \bibinfo{address}{New York, NY, USA}, Article \bibinfo{articleno}{286},
  \bibinfo{numpages}{12}~pages.
\newblock
\showISBNx{978-1-4503-5620-6}
\urldef\tempurl%
\url{https://doi.org/10.1145/3173574.3173860}
\showDOI{\tempurl}


\bibitem[\protect\citeauthoryear{Amershi, Cakmak, Knox, and Kulesza}{Amershi
  et~al\mbox{.}}{2014}]%
        {amershi_power_2014}
\bibfield{author}{\bibinfo{person}{Saleema Amershi}, \bibinfo{person}{Maya
  Cakmak}, \bibinfo{person}{William~Bradley Knox}, {and} \bibinfo{person}{Todd
  Kulesza}.} \bibinfo{year}{2014}\natexlab{}.
\newblock \showarticletitle{Power to the people: {The} role of humans in
  interactive machine learning}.
\newblock \bibinfo{journal}{\emph{AI Magazine}} \bibinfo{volume}{35},
  \bibinfo{number}{4} (\bibinfo{year}{2014}), \bibinfo{pages}{105--120}.
\newblock


\bibitem[\protect\citeauthoryear{Bojarski, Del~Testa, Dworakowski, Firner,
  Flepp, Goyal, Jackel, Monfort, Muller, Zhang, et~al\mbox{.}}{Bojarski
  et~al\mbox{.}}{2016}]%
        {bojarski2016end}
\bibfield{author}{\bibinfo{person}{Mariusz Bojarski}, \bibinfo{person}{Davide
  Del~Testa}, \bibinfo{person}{Daniel Dworakowski}, \bibinfo{person}{Bernhard
  Firner}, \bibinfo{person}{Beat Flepp}, \bibinfo{person}{Prasoon Goyal},
  \bibinfo{person}{Lawrence~D Jackel}, \bibinfo{person}{Mathew Monfort},
  \bibinfo{person}{Urs Muller}, \bibinfo{person}{Jiakai Zhang},
  {et~al\mbox{.}}} \bibinfo{year}{2016}\natexlab{}.
\newblock \showarticletitle{End to end learning for self-driving cars}.
\newblock \bibinfo{journal}{\emph{arXiv preprint arXiv:1604.07316}}
  (\bibinfo{year}{2016}).
\newblock


\bibitem[\protect\citeauthoryear{Choe, Lee, Lee, Pratt, and Kientz}{Choe
  et~al\mbox{.}}{2014}]%
        {Choe:2014:UQP:2556288.2557372}
\bibfield{author}{\bibinfo{person}{Eun~Kyoung Choe}, \bibinfo{person}{Nicole~B.
  Lee}, \bibinfo{person}{Bongshin Lee}, \bibinfo{person}{Wanda Pratt}, {and}
  \bibinfo{person}{Julie~A. Kientz}.} \bibinfo{year}{2014}\natexlab{}.
\newblock \showarticletitle{Understanding Quantified-selfers' Practices in
  Collecting and Exploring Personal Data}. In
  \bibinfo{booktitle}{\emph{Proceedings of the SIGCHI Conference on Human
  Factors in Computing Systems}} \emph{(\bibinfo{series}{CHI '14})}.
  \bibinfo{publisher}{ACM}, \bibinfo{address}{New York, NY, USA},
  \bibinfo{pages}{1143--1152}.
\newblock
\showISBNx{978-1-4503-2473-1}
\urldef\tempurl%
\url{https://doi.org/10.1145/2556288.2557372}
\showDOI{\tempurl}


\bibitem[\protect\citeauthoryear{Cramer, Evers, van Someren, and
  Wielinga}{Cramer et~al\mbox{.}}{2009}]%
        {cramer_awareness_2009}
\bibfield{author}{\bibinfo{person}{Henriette~S.M. Cramer},
  \bibinfo{person}{Vanessa Evers}, \bibinfo{person}{Maarten~W. van Someren},
  {and} \bibinfo{person}{Bob~J. Wielinga}.} \bibinfo{year}{2009}\natexlab{}.
\newblock \showarticletitle{Awareness, {Training} and {Trust} in {Interaction}
  with {Adaptive} {Spam} {Filters}}. In \bibinfo{booktitle}{\emph{Proceedings
  of the {SIGCHI} {Conference} on {Human} {Factors} in {Computing} {Systems}}}
  \emph{(\bibinfo{series}{{CHI} '09})}. \bibinfo{publisher}{ACM},
  \bibinfo{address}{New York, NY, USA}, \bibinfo{pages}{909--912}.
\newblock
\showISBNx{978-1-60558-246-7}
\urldef\tempurl%
\url{https://doi.org/10.1145/1518701.1518839}
\showDOI{\tempurl}


\bibitem[\protect\citeauthoryear{Djajadiningrat, Gaver, and
  Fres}{Djajadiningrat et~al\mbox{.}}{2000}]%
        {Djajadiningrat:2000:IRE:347642.347664}
\bibfield{author}{\bibinfo{person}{J.~P. Djajadiningrat},
  \bibinfo{person}{W.~W. Gaver}, {and} \bibinfo{person}{J.~W. Fres}.}
  \bibinfo{year}{2000}\natexlab{}.
\newblock \showarticletitle{Interaction Relabelling and Extreme Characters:
  Methods for Exploring Aesthetic Interactions}. In
  \bibinfo{booktitle}{\emph{Proceedings of the 3rd Conference on Designing
  Interactive Systems: Processes, Practices, Methods, and Techniques}}
  \emph{(\bibinfo{series}{DIS '00})}. \bibinfo{publisher}{ACM},
  \bibinfo{address}{New York, NY, USA}, \bibinfo{pages}{66--71}.
\newblock
\showISBNx{1-58113-219-0}
\urldef\tempurl%
\url{https://doi.org/10.1145/347642.347664}
\showDOI{\tempurl}


\bibitem[\protect\citeauthoryear{Dove, Halskov, Forlizzi, and Zimmerman}{Dove
  et~al\mbox{.}}{2017}]%
        {Dove:2017:UDI:3025453.3025739}
\bibfield{author}{\bibinfo{person}{Graham Dove}, \bibinfo{person}{Kim Halskov},
  \bibinfo{person}{Jodi Forlizzi}, {and} \bibinfo{person}{John Zimmerman}.}
  \bibinfo{year}{2017}\natexlab{}.
\newblock \showarticletitle{UX Design Innovation: Challenges for Working with
  Machine Learning As a Design Material}. In
  \bibinfo{booktitle}{\emph{Proceedings of the 2017 CHI Conference on Human
  Factors in Computing Systems}} \emph{(\bibinfo{series}{CHI '17})}.
  \bibinfo{publisher}{ACM}, \bibinfo{address}{New York, NY, USA},
  \bibinfo{pages}{278--288}.
\newblock
\showISBNx{978-1-4503-4655-9}
\urldef\tempurl%
\url{https://doi.org/10.1145/3025453.3025739}
\showDOI{\tempurl}


\bibitem[\protect\citeauthoryear{Eslami, Rickman, Vaccaro, Aleyasen, Vuong,
  Karahalios, Hamilton, and Sandvig}{Eslami et~al\mbox{.}}{2015}]%
        {eslami_i_2015}
\bibfield{author}{\bibinfo{person}{Motahhare Eslami}, \bibinfo{person}{Aimee
  Rickman}, \bibinfo{person}{Kristen Vaccaro}, \bibinfo{person}{Amirhossein
  Aleyasen}, \bibinfo{person}{Andy Vuong}, \bibinfo{person}{Karrie Karahalios},
  \bibinfo{person}{Kevin Hamilton}, {and} \bibinfo{person}{Christian Sandvig}.}
  \bibinfo{year}{2015}\natexlab{}.
\newblock \showarticletitle{"I Always Assumed That I Wasn'T Really That Close
  to [Her]": Reasoning About Invisible Algorithms in News Feeds}. In
  \bibinfo{booktitle}{\emph{Proceedings of the 33rd Annual ACM Conference on
  Human Factors in Computing Systems}} \emph{(\bibinfo{series}{CHI '15})}.
  \bibinfo{publisher}{ACM}, \bibinfo{address}{New York, NY, USA},
  \bibinfo{pages}{153--162}.
\newblock
\showISBNx{978-1-4503-3145-6}
\urldef\tempurl%
\url{https://doi.org/10.1145/2702123.2702556}
\showDOI{\tempurl}


\bibitem[\protect\citeauthoryear{Eslami, Vaccaro, Karahalios, and
  Hamilton}{Eslami et~al\mbox{.}}{2017}]%
        {eslami2017careful}
\bibfield{author}{\bibinfo{person}{Motahhare Eslami}, \bibinfo{person}{Kristen
  Vaccaro}, \bibinfo{person}{Karrie Karahalios}, {and} \bibinfo{person}{Kevin
  Hamilton}.} \bibinfo{year}{2017}\natexlab{}.
\newblock \showarticletitle{"Be Careful; Things Can Be Worse than They Appear":
  Understanding Biased Algorithms and Users' Behavior Around Them in Rating
  Platforms.}. In \bibinfo{booktitle}{\emph{ICWSM}}. \bibinfo{pages}{62--71}.
\newblock


\bibitem[\protect\citeauthoryear{Goodfellow, Bengio, and Courville}{Goodfellow
  et~al\mbox{.}}{2016}]%
        {Goodfellow-et-al-2016}
\bibfield{author}{\bibinfo{person}{Ian Goodfellow}, \bibinfo{person}{Yoshua
  Bengio}, {and} \bibinfo{person}{Aaron Courville}.}
  \bibinfo{year}{2016}\natexlab{}.
\newblock \bibinfo{booktitle}{\emph{Deep Learning}}.
\newblock \bibinfo{publisher}{MIT Press}.
\newblock
\newblock
\shownote{\url{http://www.deeplearningbook.org}.}


\bibitem[\protect\citeauthoryear{Graham}{Graham}{2014}]%
        {DBLP:journals/corr/Graham14a}
\bibfield{author}{\bibinfo{person}{Benjamin Graham}.}
  \bibinfo{year}{2014}\natexlab{}.
\newblock \showarticletitle{Fractional Max-Pooling}.
\newblock \bibinfo{journal}{\emph{CoRR}}  \bibinfo{volume}{abs/1412.6071}
  (\bibinfo{year}{2014}).
\newblock
\showeprint[arxiv]{1412.6071}
\urldef\tempurl%
\url{http://arxiv.org/abs/1412.6071}
\showURL{%
\tempurl}


\bibitem[\protect\citeauthoryear{Gregor and Benbasat}{Gregor and
  Benbasat}{1999}]%
        {gregor1999explanations}
\bibfield{author}{\bibinfo{person}{Shirley Gregor} {and} \bibinfo{person}{Izak
  Benbasat}.} \bibinfo{year}{1999}\natexlab{}.
\newblock \showarticletitle{Explanations from intelligent systems: Theoretical
  foundations and implications for practice}.
\newblock \bibinfo{journal}{\emph{MIS quarterly}} (\bibinfo{year}{1999}),
  \bibinfo{pages}{497--530}.
\newblock


\bibitem[\protect\citeauthoryear{Griggio, McGrenere, and Mackay}{Griggio
  et~al\mbox{.}}{2019}]%
        {griggio_customizations}
\bibfield{author}{\bibinfo{person}{Carla~F. Griggio}, \bibinfo{person}{Joanna
  McGrenere}, {and} \bibinfo{person}{Wendy Mackay}.}
  \bibinfo{year}{2019}\natexlab{}.
\newblock \showarticletitle{{Customizations and Expression Breakdowns in
  Ecosystems of Communication Apps}}. In \bibinfo{booktitle}{\emph{{CSCW
  2019}}}. \bibinfo{address}{Austin, Texas}.
\newblock


\bibitem[\protect\citeauthoryear{Hamilton, Karahalios, Sandvig, and
  Eslami}{Hamilton et~al\mbox{.}}{2014}]%
        {Hamilton:2014:PUE:2559206.2578883}
\bibfield{author}{\bibinfo{person}{Kevin Hamilton}, \bibinfo{person}{Karrie
  Karahalios}, \bibinfo{person}{Christian Sandvig}, {and}
  \bibinfo{person}{Motahhare Eslami}.} \bibinfo{year}{2014}\natexlab{}.
\newblock \showarticletitle{A Path to Understanding the Effects of Algorithm
  Awareness}. In \bibinfo{booktitle}{\emph{CHI '14 Extended Abstracts on Human
  Factors in Computing Systems}} \emph{(\bibinfo{series}{CHI EA '14})}.
  \bibinfo{publisher}{ACM}, \bibinfo{address}{New York, NY, USA},
  \bibinfo{pages}{631--642}.
\newblock
\showISBNx{978-1-4503-2474-8}
\urldef\tempurl%
\url{https://doi.org/10.1145/2559206.2578883}
\showDOI{\tempurl}


\bibitem[\protect\citeauthoryear{Hardt}{Hardt}{2014}]%
        {hardt_how_2014}
\bibfield{author}{\bibinfo{person}{Moritz Hardt}.}
  \bibinfo{year}{2014}\natexlab{}.
\newblock \bibinfo{title}{How big data is unfair}.
\newblock
\newblock
\urldef\tempurl%
\url{https://medium.com/@mrtz/how-big-data-is-unfair-9aa544d739de}
\showURL{%
\tempurl}


\bibitem[\protect\citeauthoryear{Herlocker, Konstan, and Riedl}{Herlocker
  et~al\mbox{.}}{2000}]%
        {herlocker2000explaining}
\bibfield{author}{\bibinfo{person}{Jonathan~L Herlocker},
  \bibinfo{person}{Joseph~A Konstan}, {and} \bibinfo{person}{John Riedl}.}
  \bibinfo{year}{2000}\natexlab{}.
\newblock \showarticletitle{Explaining collaborative filtering
  recommendations}. In \bibinfo{booktitle}{\emph{Proceedings of the 2000 ACM
  conference on Computer supported cooperative work}}. ACM,
  \bibinfo{pages}{241--250}.
\newblock


\bibitem[\protect\citeauthoryear{Hern}{Hern}{2015}]%
        {hern_flickr_2015}
\bibfield{author}{\bibinfo{person}{Alex Hern}.}
  \bibinfo{year}{2015}\natexlab{}.
\newblock \showarticletitle{Flickr faces complaints over 'offensive'
  auto-tagging for photos}.
\newblock \bibinfo{journal}{\emph{The Guardian}} (\bibinfo{date}{May}
  \bibinfo{year}{2015}).
\newblock
\showISSN{0261-3077}
\urldef\tempurl%
\url{https://www.theguardian.com/technology/2015/may/20/flickr-complaints-offensive-auto-tagging-photos}
\showURL{%
\tempurl}


\bibitem[\protect\citeauthoryear{Kim}{Kim}{2015}]%
        {kim2015interactive}
\bibfield{author}{\bibinfo{person}{Been Kim}.} \bibinfo{year}{2015}\natexlab{}.
\newblock \emph{\bibinfo{title}{Interactive and interpretable machine learning
  models for human machine collaboration}}.
\newblock \bibinfo{thesistype}{Ph.D. Dissertation}.
  \bibinfo{school}{Massachusetts Institute of Technology}.
\newblock


\bibitem[\protect\citeauthoryear{Krizhevsky}{Krizhevsky}{2009}]%
        {krizhevsky2009learning}
\bibfield{author}{\bibinfo{person}{Alex Krizhevsky}.}
  \bibinfo{year}{2009}\natexlab{}.
\newblock \showarticletitle{Learning multiple layers of features from tiny
  images}.
\newblock  (\bibinfo{year}{2009}).
\newblock


\bibitem[\protect\citeauthoryear{Lee and See}{Lee and See}{2004}]%
        {lee2004trust}
\bibfield{author}{\bibinfo{person}{John~D Lee} {and} \bibinfo{person}{Katrina~A
  See}.} \bibinfo{year}{2004}\natexlab{}.
\newblock \showarticletitle{Trust in automation: Designing for appropriate
  reliance}.
\newblock \bibinfo{journal}{\emph{Human Factors: The Journal of the Human
  Factors and Ergonomics Society}} \bibinfo{volume}{46}, \bibinfo{number}{1}
  (\bibinfo{year}{2004}), \bibinfo{pages}{50--80}.
\newblock


\bibitem[\protect\citeauthoryear{Liang and Hu}{Liang and Hu}{2015}]%
        {liang2015recurrent}
\bibfield{author}{\bibinfo{person}{Ming Liang} {and} \bibinfo{person}{Xiaolin
  Hu}.} \bibinfo{year}{2015}\natexlab{}.
\newblock \showarticletitle{Recurrent convolutional neural network for object
  recognition}. In \bibinfo{booktitle}{\emph{Proceedings of the IEEE Conference
  on Computer Vision and Pattern Recognition}}. \bibinfo{pages}{3367--3375}.
\newblock


\bibitem[\protect\citeauthoryear{Luhmann}{Luhmann}{1979}]%
        {luhmann_trust_1979}
\bibfield{author}{\bibinfo{person}{Niklas Luhmann}.}
  \bibinfo{year}{1979}\natexlab{}.
\newblock \showarticletitle{Trust and power. 1979}.
\newblock \bibinfo{journal}{\emph{John Willey \& Sons}} (\bibinfo{year}{1979}).
\newblock


\bibitem[\protect\citeauthoryear{MacLeod, Bennett, Morris, and Cutrell}{MacLeod
  et~al\mbox{.}}{2017}]%
        {MacLeod:2017:UBP:3025453.3025814}
\bibfield{author}{\bibinfo{person}{Haley MacLeod}, \bibinfo{person}{Cynthia~L.
  Bennett}, \bibinfo{person}{Meredith~Ringel Morris}, {and}
  \bibinfo{person}{Edward Cutrell}.} \bibinfo{year}{2017}\natexlab{}.
\newblock \showarticletitle{Understanding Blind People's Experiences with
  Computer-Generated Captions of Social Media Images}. In
  \bibinfo{booktitle}{\emph{Proceedings of the 2017 CHI Conference on Human
  Factors in Computing Systems}} \emph{(\bibinfo{series}{CHI '17})}.
  \bibinfo{publisher}{ACM}, \bibinfo{address}{New York, NY, USA},
  \bibinfo{pages}{5988--5999}.
\newblock
\showISBNx{978-1-4503-4655-9}
\urldef\tempurl%
\url{https://doi.org/10.1145/3025453.3025814}
\showDOI{\tempurl}


\bibitem[\protect\citeauthoryear{Mayring}{Mayring}{2007}]%
        {mayring2007generalization}
\bibfield{author}{\bibinfo{person}{Philipp Mayring}.}
  \bibinfo{year}{2007}\natexlab{}.
\newblock \showarticletitle{On generalization in qualitatively oriented
  research}. In \bibinfo{booktitle}{\emph{Forum Qualitative
  Sozialforschung/Forum: Qualitative Social Research}},
  Vol.~\bibinfo{volume}{8}.
\newblock


\bibitem[\protect\citeauthoryear{Mishkin and Matas}{Mishkin and Matas}{2015}]%
        {DBLP:journals/corr/MishkinM15}
\bibfield{author}{\bibinfo{person}{Dmytro Mishkin} {and} \bibinfo{person}{Jiri
  Matas}.} \bibinfo{year}{2015}\natexlab{}.
\newblock \showarticletitle{All you need is a good init}.
\newblock \bibinfo{journal}{\emph{CoRR}}  \bibinfo{volume}{abs/1511.06422}
  (\bibinfo{year}{2015}).
\newblock
\showeprint[arxiv]{1511.06422}
\urldef\tempurl%
\url{http://arxiv.org/abs/1511.06422}
\showURL{%
\tempurl}


\bibitem[\protect\citeauthoryear{Muir}{Muir}{1994}]%
        {muir_trust_1994}
\bibfield{author}{\bibinfo{person}{Bonnie~M. Muir}.}
  \bibinfo{year}{1994}\natexlab{}.
\newblock \showarticletitle{Trust in automation: {Part} {I}. {Theoretical}
  issues in the study of trust and human intervention in automated systems}.
\newblock \bibinfo{journal}{\emph{Ergonomics}} \bibinfo{volume}{37},
  \bibinfo{number}{11} (\bibinfo{date}{Nov.} \bibinfo{year}{1994}),
  \bibinfo{pages}{1905--1922}.
\newblock
\showISSN{0014-0139}
\urldef\tempurl%
\url{https://doi.org/10.1080/00140139408964957}
\showDOI{\tempurl}


\bibitem[\protect\citeauthoryear{Muir and Moray}{Muir and Moray}{1996}]%
        {muir_trust_1996}
\bibfield{author}{\bibinfo{person}{B.~M. Muir} {and} \bibinfo{person}{N.
  Moray}.} \bibinfo{year}{1996}\natexlab{}.
\newblock \showarticletitle{Trust in automation. {Part} {II}. {Experimental}
  studies of trust and human intervention in a process control simulation}.
\newblock \bibinfo{journal}{\emph{Ergonomics}} \bibinfo{volume}{39},
  \bibinfo{number}{3} (\bibinfo{date}{March} \bibinfo{year}{1996}),
  \bibinfo{pages}{429--460}.
\newblock
\showISSN{0014-0139}
\urldef\tempurl%
\url{https://doi.org/10.1080/00140139608964474}
\showDOI{\tempurl}


\bibitem[\protect\citeauthoryear{Qin, Yu, Liu, and Chen}{Qin
  et~al\mbox{.}}{2018}]%
        {DBLP:journals/corr/abs-1804-11191}
\bibfield{author}{\bibinfo{person}{Zhuwei Qin}, \bibinfo{person}{Fuxun Yu},
  \bibinfo{person}{Chenchen Liu}, {and} \bibinfo{person}{Xiang Chen}.}
  \bibinfo{year}{2018}\natexlab{}.
\newblock \showarticletitle{How convolutional neural network see the world -
  {A} survey of convolutional neural network visualization methods}.
\newblock \bibinfo{journal}{\emph{CoRR}}  \bibinfo{volume}{abs/1804.11191}
  (\bibinfo{year}{2018}).
\newblock
\showeprint[arxiv]{1804.11191}
\urldef\tempurl%
\url{http://arxiv.org/abs/1804.11191}
\showURL{%
\tempurl}


\bibitem[\protect\citeauthoryear{Rader, Cotter, and Cho}{Rader
  et~al\mbox{.}}{2018}]%
        {Rader:2018:EMS:3173574.3173677}
\bibfield{author}{\bibinfo{person}{Emilee Rader}, \bibinfo{person}{Kelley
  Cotter}, {and} \bibinfo{person}{Janghee Cho}.}
  \bibinfo{year}{2018}\natexlab{}.
\newblock \showarticletitle{Explanations As Mechanisms for Supporting
  Algorithmic Transparency}. In \bibinfo{booktitle}{\emph{Proceedings of the
  2018 CHI Conference on Human Factors in Computing Systems}}
  \emph{(\bibinfo{series}{CHI '18})}. \bibinfo{publisher}{ACM},
  \bibinfo{address}{New York, NY, USA}, Article \bibinfo{articleno}{103},
  \bibinfo{numpages}{13}~pages.
\newblock
\showISBNx{978-1-4503-5620-6}
\urldef\tempurl%
\url{https://doi.org/10.1145/3173574.3173677}
\showDOI{\tempurl}


\bibitem[\protect\citeauthoryear{Rauber, Fadel, Falcão, and Telea}{Rauber
  et~al\mbox{.}}{2017}]%
        {7539329}
\bibfield{author}{\bibinfo{person}{P.~E. Rauber}, \bibinfo{person}{S.~G.
  Fadel}, \bibinfo{person}{A.~X. Falcão}, {and} \bibinfo{person}{A.~C.
  Telea}.} \bibinfo{year}{2017}\natexlab{}.
\newblock \showarticletitle{Visualizing the Hidden Activity of Artificial
  Neural Networks}.
\newblock \bibinfo{journal}{\emph{IEEE Transactions on Visualization and
  Computer Graphics}} \bibinfo{volume}{23}, \bibinfo{number}{1}
  (\bibinfo{date}{Jan} \bibinfo{year}{2017}), \bibinfo{pages}{101--110}.
\newblock
\showISSN{1077-2626}
\urldef\tempurl%
\url{https://doi.org/10.1109/TVCG.2016.2598838}
\showDOI{\tempurl}


\bibitem[\protect\citeauthoryear{Ribeiro, Singh, and Guestrin}{Ribeiro
  et~al\mbox{.}}{2016}]%
        {Ribeiro_2016_WIT}
\bibfield{author}{\bibinfo{person}{Marco~Tulio Ribeiro},
  \bibinfo{person}{Sameer Singh}, {and} \bibinfo{person}{Carlos Guestrin}.}
  \bibinfo{year}{2016}\natexlab{}.
\newblock \showarticletitle{"Why Should I Trust You?": Explaining the
  Predictions of Any Classifier}. In \bibinfo{booktitle}{\emph{Proceedings of
  the 22Nd ACM SIGKDD International Conference on Knowledge Discovery and Data
  Mining}} \emph{(\bibinfo{series}{KDD '16})}. \bibinfo{publisher}{ACM},
  \bibinfo{address}{New York, NY, USA}, \bibinfo{pages}{1135--1144}.
\newblock
\showISBNx{978-1-4503-4232-2}
\urldef\tempurl%
\url{https://doi.org/10.1145/2939672.2939778}
\showDOI{\tempurl}


\bibitem[\protect\citeauthoryear{Rousseau, Sitkin, Burt, and Camerer}{Rousseau
  et~al\mbox{.}}{1998}]%
        {rousseau_not_1998}
\bibfield{author}{\bibinfo{person}{Denise~M. Rousseau}, \bibinfo{person}{Sim~B.
  Sitkin}, \bibinfo{person}{Ronald~S. Burt}, {and} \bibinfo{person}{Colin
  Camerer}.} \bibinfo{year}{1998}\natexlab{}.
\newblock \showarticletitle{Not {So} {Different} {After} {All}: {A}
  {Cross}-{Discipline} {View} {Of} {Trust}}.
\newblock \bibinfo{journal}{\emph{Academy of Management Review}}
  \bibinfo{volume}{23}, \bibinfo{number}{3} (\bibinfo{date}{July}
  \bibinfo{year}{1998}), \bibinfo{pages}{393--404}.
\newblock
\showISSN{0363-7425, 1930-3807}
\urldef\tempurl%
\url{https://doi.org/10.5465/AMR.1998.926617}
\showDOI{\tempurl}


\bibitem[\protect\citeauthoryear{Sabharwal}{Sabharwal}{2015}]%
        {google_photos_2015}
\bibfield{author}{\bibinfo{person}{Anil Sabharwal}.}
  \bibinfo{year}{2015}\natexlab{}.
\newblock \bibinfo{title}{Picture this: {A} fresh approach to {Photos}}.
\newblock
\newblock
\urldef\tempurl%
\url{https://blog.google/products/photos/picture-this-fresh-approach-to-photos/}
\showURL{%
\tempurl}


\bibitem[\protect\citeauthoryear{Samek, Binder, Montavon, Bach, and
  M{\"{u}}ller}{Samek et~al\mbox{.}}{2015}]%
        {DBLP:journals/corr/SamekBMBM15}
\bibfield{author}{\bibinfo{person}{Wojciech Samek}, \bibinfo{person}{Alexander
  Binder}, \bibinfo{person}{Gr{\'{e}}goire Montavon},
  \bibinfo{person}{Sebastian Bach}, {and} \bibinfo{person}{Klaus{-}Robert
  M{\"{u}}ller}.} \bibinfo{year}{2015}\natexlab{}.
\newblock \showarticletitle{Evaluating the visualization of what a Deep Neural
  Network has learned}.
\newblock \bibinfo{journal}{\emph{CoRR}}  \bibinfo{volume}{abs/1509.06321}
  (\bibinfo{year}{2015}).
\newblock
\showeprint[arxiv]{1509.06321}
\urldef\tempurl%
\url{http://arxiv.org/abs/1509.06321}
\showURL{%
\tempurl}


\bibitem[\protect\citeauthoryear{Selvaraju, Das, Vedantam, Cogswell, Parikh,
  and Batra}{Selvaraju et~al\mbox{.}}{2016}]%
        {DBLP:journals/corr/SelvarajuDVCPB16}
\bibfield{author}{\bibinfo{person}{Ramprasaath~R. Selvaraju},
  \bibinfo{person}{Abhishek Das}, \bibinfo{person}{Ramakrishna Vedantam},
  \bibinfo{person}{Michael Cogswell}, \bibinfo{person}{Devi Parikh}, {and}
  \bibinfo{person}{Dhruv Batra}.} \bibinfo{year}{2016}\natexlab{}.
\newblock \showarticletitle{Grad-CAM: Why did you say that? Visual Explanations
  from Deep Networks via Gradient-based Localization}.
\newblock \bibinfo{journal}{\emph{CoRR}}  \bibinfo{volume}{abs/1610.02391}
  (\bibinfo{year}{2016}).
\newblock
\showeprint[arxiv]{1610.02391}
\urldef\tempurl%
\url{http://arxiv.org/abs/1610.02391}
\showURL{%
\tempurl}


\bibitem[\protect\citeauthoryear{Springenberg, Dosovitskiy, Brox, and
  Riedmiller}{Springenberg et~al\mbox{.}}{2014}]%
        {DBLP:journals/corr/SpringenbergDBR14}
\bibfield{author}{\bibinfo{person}{Jost~Tobias Springenberg},
  \bibinfo{person}{Alexey Dosovitskiy}, \bibinfo{person}{Thomas Brox}, {and}
  \bibinfo{person}{Martin~A. Riedmiller}.} \bibinfo{year}{2014}\natexlab{}.
\newblock \showarticletitle{Striving for Simplicity: The All Convolutional
  Net}.
\newblock \bibinfo{journal}{\emph{CoRR}}  \bibinfo{volume}{abs/1412.6806}
  (\bibinfo{year}{2014}).
\newblock
\urldef\tempurl%
\url{http://arxiv.org/abs/1412.6806}
\showURL{%
\tempurl}


\bibitem[\protect\citeauthoryear{Strobelt, Gehrmann, Huber, Pfister, and
  Rush}{Strobelt et~al\mbox{.}}{2016}]%
        {DBLP:journals/corr/StrobeltGHPR16}
\bibfield{author}{\bibinfo{person}{Hendrik Strobelt},
  \bibinfo{person}{Sebastian Gehrmann}, \bibinfo{person}{Bernd Huber},
  \bibinfo{person}{Hanspeter Pfister}, {and} \bibinfo{person}{Alexander~M.
  Rush}.} \bibinfo{year}{2016}\natexlab{}.
\newblock \showarticletitle{Visual Analysis of Hidden State Dynamics in
  Recurrent Neural Networks}.
\newblock \bibinfo{journal}{\emph{CoRR}}  \bibinfo{volume}{abs/1606.07461}
  (\bibinfo{year}{2016}).
\newblock
\showeprint[arxiv]{1606.07461}
\urldef\tempurl%
\url{http://arxiv.org/abs/1606.07461}
\showURL{%
\tempurl}


\bibitem[\protect\citeauthoryear{Stumpf, Rajaram, Li, Wong, Burnett,
  Dietterich, Sullivan, and Herlocker}{Stumpf et~al\mbox{.}}{2009}]%
        {STUMPF2009639}
\bibfield{author}{\bibinfo{person}{Simone Stumpf}, \bibinfo{person}{Vidya
  Rajaram}, \bibinfo{person}{Lida Li}, \bibinfo{person}{Weng-Keen Wong},
  \bibinfo{person}{Margaret Burnett}, \bibinfo{person}{Thomas Dietterich},
  \bibinfo{person}{Erin Sullivan}, {and} \bibinfo{person}{Jonathan Herlocker}.}
  \bibinfo{year}{2009}\natexlab{}.
\newblock \showarticletitle{Interacting meaningfully with machine learning
  systems: Three experiments}.
\newblock \bibinfo{journal}{\emph{International Journal of Human-Computer
  Studies}} \bibinfo{volume}{67}, \bibinfo{number}{8} (\bibinfo{year}{2009}),
  \bibinfo{pages}{639 -- 662}.
\newblock
\showISSN{1071-5819}
\urldef\tempurl%
\url{https://doi.org/10.1016/j.ijhcs.2009.03.004}
\showDOI{\tempurl}


\bibitem[\protect\citeauthoryear{Teichman and Thrun}{Teichman and
  Thrun}{2011}]%
        {6301978}
\bibfield{author}{\bibinfo{person}{A. Teichman} {and} \bibinfo{person}{S.
  Thrun}.} \bibinfo{year}{2011}\natexlab{}.
\newblock \showarticletitle{Practical object recognition in autonomous driving
  and beyond}. In \bibinfo{booktitle}{\emph{Advanced Robotics and its Social
  Impacts}}. \bibinfo{pages}{35--38}.
\newblock
\showISSN{2162-7576}
\urldef\tempurl%
\url{https://doi.org/10.1109/ARSO.2011.6301978}
\showDOI{\tempurl}


\bibitem[\protect\citeauthoryear{Tufte}{Tufte}{1986}]%
        {Tufte:1986:VDQ:33404}
\bibfield{author}{\bibinfo{person}{Edward~R. Tufte}.}
  \bibinfo{year}{1986}\natexlab{}.
\newblock \bibinfo{booktitle}{\emph{The Visual Display of Quantitative
  Information}}.
\newblock \bibinfo{publisher}{Graphics Press}, \bibinfo{address}{Cheshire, CT,
  USA}.
\newblock
\showISBNx{0-9613921-0-X}


\bibitem[\protect\citeauthoryear{Tullio, Dey, Chalecki, and Fogarty}{Tullio
  et~al\mbox{.}}{2007}]%
        {tullio2007works}
\bibfield{author}{\bibinfo{person}{Joe Tullio}, \bibinfo{person}{Anind~K Dey},
  \bibinfo{person}{Jason Chalecki}, {and} \bibinfo{person}{James Fogarty}.}
  \bibinfo{year}{2007}\natexlab{}.
\newblock \showarticletitle{How it works: a field study of non-technical users
  interacting with an intelligent system}. In
  \bibinfo{booktitle}{\emph{Proceedings of the SIGCHI Conference on Human
  Factors in Computing Systems}}. ACM, \bibinfo{pages}{31--40}.
\newblock


\bibitem[\protect\citeauthoryear{Veale, Van~Kleek, and Binns}{Veale
  et~al\mbox{.}}{2018}]%
        {Veale:2018:FAD:3173574.3174014}
\bibfield{author}{\bibinfo{person}{Michael Veale}, \bibinfo{person}{Max
  Van~Kleek}, {and} \bibinfo{person}{Reuben Binns}.}
  \bibinfo{year}{2018}\natexlab{}.
\newblock \showarticletitle{Fairness and Accountability Design Needs for
  Algorithmic Support in High-Stakes Public Sector Decision-Making}. In
  \bibinfo{booktitle}{\emph{Proceedings of the 2018 CHI Conference on Human
  Factors in Computing Systems}} \emph{(\bibinfo{series}{CHI '18})}.
  \bibinfo{publisher}{ACM}, \bibinfo{address}{New York, NY, USA}, Article
  \bibinfo{articleno}{440}, \bibinfo{numpages}{14}~pages.
\newblock
\showISBNx{978-1-4503-5620-6}
\urldef\tempurl%
\url{https://doi.org/10.1145/3173574.3174014}
\showDOI{\tempurl}


\bibitem[\protect\citeauthoryear{Yosinski, Clune, Nguyen, Fuchs, and
  Lipson}{Yosinski et~al\mbox{.}}{2015}]%
        {DBLP:journals/corr/YosinskiCNFL15}
\bibfield{author}{\bibinfo{person}{Jason Yosinski}, \bibinfo{person}{Jeff
  Clune}, \bibinfo{person}{Anh~Mai Nguyen}, \bibinfo{person}{Thomas~J. Fuchs},
  {and} \bibinfo{person}{Hod Lipson}.} \bibinfo{year}{2015}\natexlab{}.
\newblock \showarticletitle{Understanding Neural Networks Through Deep
  Visualization}.
\newblock \bibinfo{journal}{\emph{CoRR}}  \bibinfo{volume}{abs/1506.06579}
  (\bibinfo{year}{2015}).
\newblock
\showeprint[arxiv]{1506.06579}
\urldef\tempurl%
\url{http://arxiv.org/abs/1506.06579}
\showURL{%
\tempurl}


\end{thebibliography}

\end{document}